\documentclass[fleqn,usenatbib]{mnras}
\usepackage{newtxtext,newtxmath}
\usepackage[T1]{fontenc}
\DeclareRobustCommand{\VAN}[3]{#2}
\let\VANthebibliography\thebibliography
\def\thebibliography{\DeclareRobustCommand{\VAN}[3]{##3}\VANthebibliography}
\usepackage{graphicx}	
\usepackage{amsmath}
\usepackage{aas-macros}
\usepackage[nopatch]{microtype}
\usepackage{booktabs}
\def\fdg{\hbox{$.\!\!^\circ$}}
\def\deg{\hbox{$^\circ$}}
\def\uh{\hbox{$^{\rm h}$}}
\def\um{\hbox{$^{\rm m}$}}
\def\us{\hbox{$^{\rm s}$}}
\title{Pushchino multibeam pulsar search — V. The bright FRB\,20190203 detected at 111 MHz}
\author[S. A. Tyul’bashev et al.]{S. A. Tyul’bashev$^{1}$\thanks{E-mail: serg@prao.ru (SAT)},
V. A. Samodurov$^{2,1}$,
A. S. Pozanenko$^{3,4,5}$,
E. A. Brylyakova$^{1}$,
\newauthor{
S. A. Grebenev$^{3}$,
I. V. Chelovekov$^{3}$,
P. Yu. Minaev$^{3}$,
E. A. Isaev$^{6}$,
M. V. Barkov$^{7}$,}
\\
$^{1}$	P.N. Lebedev Physical Institute, Russian Academy of Sciences, Astro Space Center, Pushchino Radio Astronomy Observatory, \\ Radiotelescopnaya 1, Moscow region, Pushchino, 142290, Russia\\
$^{2}$ Graduate School of Business, HSE University\\
$^{3}$ Space Research Institute, Russian Academy of Sciences, Profsoyusnaya 84/32, Moscow, 117997, Russia\\
$^{4}$ National Research University Higher School of Economics, Pokrovskii bulvar 11, Moscow, 109028, Russia\\
$^{5}$ Institute of Physics and Technology, Institutskiy pereulok 9, Dolgoprudny, 141701, Russia\\
$^{6}$ Institute of Mathematical Problems of Biology, Russian Academy of Sciences, Branch of the Keldysh Institute of Applied Mathematics, \\ Vitkevich 1a, Moscow region, Pushchino, 142290, Russia\\
$^{7}$ Institute of Astronomy, Russian Academy of Sciences, Pyatnitskaya 48, Moscow, 119017, Russia
}

\date{Accepted XXX. Received YYY; in original form ZZZ}

\pubyear{}

\begin{document}
	\label{firstpage}
	\pagerange{\pageref{firstpage}--\pageref{lastpage}}
	\maketitle
	
	\begin{abstract}
Since August 2014, a monitoring survey at a frequency of 111 MHz has been conducted on the Large Phased Array (LPA) radio telescope of the P.N. Lebedev Physical Institute (LPI). We report the discovery of a bright pulse having a dispersion measure ($DM$) equal to $134.4\pm2\ \mbox{pc cm}^{-3}$, a peak flux density ($S_p$) equal to $20\pm4$ Jy  and a half-width ($W_e$) equal to $211\pm6$ ms. The excessive $DM$ of the pulse, after taking into account the Milky Way contribution, is $114\ \mbox{pc cm}^{-3}$ that indicates its extragalactic origin. Such value of $DM$ corresponds to the luminosity distance 713 Mpc. The above parameters make the pulse to be a reliable candidate to the fast radio burst (FRB) event, and then it is the second FRB detected at such a large $\lambda\sim2.7$ m wavelength and the first one among non-repeating FRBs. The normalized luminosity $L_\nu$ of the event, which we have designated as FRB\,20190203, estimated under assumption that the whole excessive $DM$ is determined by the intergalactic environment toward the host galaxy, is equal to $\simeq 10^{34}\ \mbox{erg s}^{-1} \mbox{Hz}^{-1}$. In addition to the study of radio data we analyzed data from the quasi-simultaneous observations of the sky in the high energy ($\geq 80$ keV) band by the omnidirectional detector SPI/ACS aboard the INTEGRAL orbital observatory (in order to look for a possible gamma-ray counterpart of FRB\,20190203). We did not detect  any transient events exceeding the background at a statistically significant level. In the INTEGRAL archive, the FRB\,20190203 localization region has been observed many times with a total exposure of       $\sim 73.2$ days. We have analyzed the data but were unable to find any reliable short gamma-ray bursts from the FRB\,20190203 position. Finally we note that the observed properties of FRB\,20190203 can be reproduced well in the framework of a maser synchrotron model operating in the far reverse shock (at a distance of $\sim 10^{15}$ cm) of a magnetar. However, triggering the burst requires a high conversion efficiency (at the level of 1\%) of the shock wave energy into the radio emission. 
\end{abstract}

\section{Introduction}

Fast radio bursts (FRBs), which are pulsed sources of intense radio emission of extragalactic origin, were discovered in 2007 in the archive data of the 64-meter telescope in Parkes (\citeauthor{Lorimer2007}, \citeyear{Lorimer2007}). Almost all FRBs were detected at frequencies between 0.4 and 1.6 GHz, that is, in the decimeter wavelength range 
(see catalogs\footnote{www.frbcat.org/}$^,$\footnote{www.chime-frb.ca/catalog} in \citeauthor
{Petroff2016}, \citeyear{Petroff2016}; \citeauthor{CHIME2021}, \citeyear{CHIME2021}). Externally, the bursts are similar to the pulsar pulses. They are short in duration and dispersed, that is, they come first at a high frequency and then at a low one. The median value of the observed duration of the bursts, according to the FRB catalog by \citeauthor{Petroff2016} (\citeyear{Petroff2016}), gets to $\sim 2$ ms. However, there were structures in bursts observed with a duration of less than 0.1 $\mu$s (\citeauthor{Majid2021}, \citeyear{Majid2021}). 

The pulse signal, passing through the interstellar medium in the host galaxy, the intergalactic medium, and the interstellar medium in the Milky Way, experiences scattering. Its observed duration becomes longer. The pulses with high signal-to-noise (S/N) ratios have often the characteristic FRED shape with a fast rise (stepped front) and an exponential decay. 

Scattering allows us to study properties of the medium between the radiation source and us, but at the same time, scattering leads to the pulse profile smearing and decreasing in the observed S/N value of the pulse. If we are talking about observations at a certain frequency, then the further away the pulse source is, the greater the observed dispersion measure ($DM$) will be and the longer the delay due to scattering ($\tau_{sc}$). If we are talking about observations at different frequencies, then the lower the frequency of observations ($\nu$), the longer the characteristic time delay due to scattering ($\tau_{sc} \sim \nu^{-4}$) should be (\citeauthor{Lorimer2004}, \citeyear{Lorimer2004}). With the difference in frequency of observations only twice, the characteristic scattering time $\tau_{sc}$ differs by more than an order of magnitude ($\sim 2^4=16$). The scattering effect cannot be compensated for and therefore leads to a deterioration in sensitivity when searching for FRBs.

In the meter wavelength range ($\nu<300$ MHz), there were three attempts made to the moment of the full-fledged search for FRBs using new radio telescopes with wide reception bands. The Low Frequency Array 
(LOFAR; \citeauthor{vanHaarlem2013}, \citeyear{vanHaarlem2013}) and Murchison Widefield Array 
(MWA; \citeauthor{Lonsdale2009}, \citeyear{Lonsdale2009}) radio telescopes have been used for the search for FRBs at 45 and 182 MHz (\citeauthor{Coenen2014}, \citeyear{Coenen2014}; \citeauthor{Karastergiou2015}, \citeyear{Karastergiou2015}; \citeauthor{Rowlinson2016}, \citeyear{Rowlinson2016}). No bursts have been detected that gave upper limits of  the expected number of FRBs. The problems with previous search for FRBs at 111 MHz using the Large Phased Array radio telescope of P.N. Lebedev Physical Institute 
(LPA LPI; \citeauthor{Shishov2016}, \citeyear{Shishov2016}) have been described by \citeauthor{Brylyakova2023} (\citeyear{Brylyakova2023}). 

There were also a number of attempts to observe repeated FRBs, registered from different sources at high frequencies, simultaneously in the decimeter (300--800 MHz) and meter (110--180 MHz) ranges (\citeauthor{Law2017}, \citeyear{Law2017}; \citeauthor{Sokolowski2018}, \citeyear{Sokolowski2018}; \citeauthor{Houben2019}, \citeyear{Houben2019}). The bursts were recorded in both the ranges (\citeauthor{Chawla2020}, \citeyear{Chawla2020}; \citeauthor{Pleunis2021}, \citeyear{Pleunis2021}) from only one unique source FRB\,20180916 (\citeauthor{Marcote2020}, \citeyear{Marcote2020}) which exhibits periodic activity. And even in this case there was no simultaneous registration for any bursts. Bursts found in decimeter range were not detected in meter range. Moreover it was noted that the burst activity was systematically delayed toward lower frequencies by about 3 days from 600 to 150 MHz. 

Since the sensitivity of observations decreases due to scattering $\sim \tau_{sc} ^{1/2}$, the smaller the distance to the burst source and, consequently, the smaller the scattering, the weaker bursts can be observed with all other parameters unchanged. The higher the fluctuation sensitivity of the radio telescope, which is proportional to the effective area, the more chances we have to register a burst. At the present time, the most sensitive meter-range radio telescopes in the world are LOFAR, MWA, LPA. The meter range is also claimed for Five-hundred-meter Aperture Spherical Telescope (FAST; \citeauthor{Nan2011}, \citeyear{Nan2011}) which should have record sensitivity in both the decimeter and meter ranges. However we were unable to find any publications based on the FAST data with attempts to detect FRBs on meter waves. In this paper, we describe our search for dispersed signals in metadata obtained during technical assessment of the quality of observations carried out with the LPA LPI radio telescope.

\section{LPA observations and data processing}

The meridian instrument LPA LPI is a phased array consisting of 16,384 dipoles. The input signal coming from the lines on which 128 dipoles are located goes to the amplifiers. The amplifiers have one input and four outputs. Therefore, there is a technical possibility to create four independent radio telescopes on the basis of a single antenna field (\citeauthor{Shishov2016}, \citeyear{Shishov2016}). Currently, three outputs are involved, one of which is used in the monitoring program. This radio telescope is called LPA3. It has 128 uncontrolled (stationary) beams overlapping declinations $-9\deg<\delta<+55\deg$. The beams are created using the Butler matrix and provide overlap at the level of 0.405. The beam size is approximately $0\fdg5\times 1\fdg0$, and its shape obeys the dependence $\sin^2(x)/x^2$. The central frequency of the observations is 110.25 MHz, the receiving frequency band is 2.5 MHz. The signal on the recorder is split into two ones and recorded onto the hard disk twice with different time-frequency resolutions. The data recorded in 6 frequency channels with a channel width of 415 kHz and a sampling time of 0.1~s are used in the ''Space Weather'' project (\citeauthor{Shishov2016}, \citeyear{Shishov2016}). Data recorded in 32 frequency channels with a channel width of 78 kHz and a sampling time of 12.5ms are used to search for pulsars and transients in the project Pushchino Multibeams Pulsar Search (PUMPS; \citeauthor{Tyulbashev2016}, \citeyear{Tyulbashev2016}; \citeauthor{Tyulbashev2018b}, \citeyear{Tyulbashev2018b}; \citeauthor{Tyulbashev2022}, \citeyear{Tyulbashev2022}). Data are recorded at one-hour intervals. The beginning of the recording is controlled by the atomic frequency standard. Inside the one-hour recording, the time is controlled by a quartz oscillator.  Six times a day, a signal of a known temperature (calibration step) is transmitted to the input amplifiers. Using this step, the data in the frequency channels can be equalized in gain.

 Quality control of observations is carried out by a special processing program. This program equalizes the gain in frequency channels, cuts all the data inside the one-hour interval into 10-s long segments (800 points in the frequency channel, $800\times 32=25,600$ points in all channels) and prepares metadata which are written into files. On each segment and in each frequency channel, the baseline is subtracted and after that the parameters (metadata) are evaluated. The metadata includes the maximum and minimum values on a segment, the average and median average values on a segment, estimates of the root mean square (RMS) deviation of noise, and others. In addition, the start time of the segment and the number of the maximum point inside a segment for each frequency channel are stored (for more details, see \citeauthor{Samodurov2022}, \citeyear{Samodurov2022}). If the maxima are detected inside the frequency channels at a given S/N, it is possible to estimate the time shift of these maxima between channels due to the dispersion delay and identify candidates for extraterrestrial origin. If the maxima in the frequency channels fall on one point, then there is no dispersion delay and, most likely, there is interference.

The volume of metadata recorded in the files is about 100 times smaller than the volume of input data. Further work with files is carried out using a program that, in accordance with a set of specified criteria, identifies interferences and sources of unknown nature. In total 153,729 one-hour files were processed to obtain metadata (or 17,000 square degrees every day in the interval of 3,000 days.) Three main criteria were laid down in the search program for dispersed pulses. Firstly, when passing from a high frequency to a low one, there should be a pulse shift in the frequency channels corresponding to $DM >100$ pc/cm$^3$. Secondly, in the tested segment there should be at least 4 frequency channels in which the signal has S/N$>5$. For 32-channel data, the RMS deviations of noise ($\sigma_n$) in one frequency channel are approximately 1.5 Jy outside the Galactic plane in the direction to the zenith. That is, only those sources were selected whose peak flux density ($S_{p}$ ), after compensation for the assumed value of $DM$, was higher than 7.5 Jy. Thirdly, the signal should be observed in no more than two adjacent beams. A special study of interference on LPA3 with the use of metadata showed that interference is observed in three or more beams, and most often in all beams of the radio telescope simultaneously (see Figure~1 in \citeauthor{Samodurov2022}, \citeyear{Samodurov2022}).

If necessary, further investigation of the sources is carried out using the original (raw) data that are stored on the server. Despite a small set of output parameters, it was based on the metadata that the first rotating radio transients (RRAT) were detected on LPA3 (\citeauthor{Samodurov2017}, \citeyear{Samodurov2017}). 
\begin{figure}
	\includegraphics[width=\columnwidth]{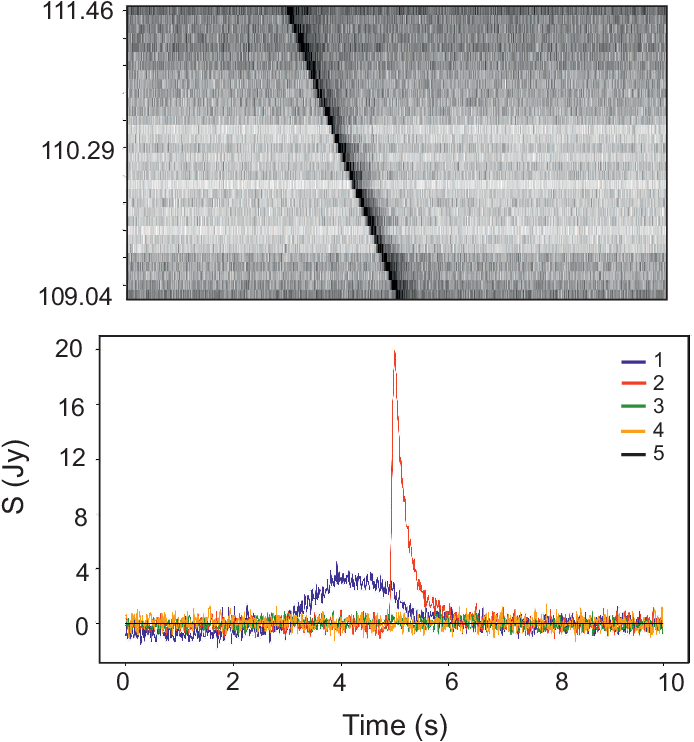}
	\caption{The upper panel shows the dynamic spectrum of the detected pulse. Frequency channels (in MHz) are indicated along the vertical axis. The lower panel shows the signals received for: 1) frequency channels summed up with $DM = 0\ \mbox{pc cm}^{-3}$; 2) channels summed up with $DM = 134.4\ \mbox{pc cm}^{-3}$; 3-4)  channels summed up with $DM = 0\ \mbox{pc cm}^{-3}$ for beams above and below the detected burst; 5) base line. The lower panel along the vertical axis shows the flux density. The horizontal axis is the same for both panels and shows the time.}
    \label{fig:fig1} 
\end{figure}

\section{Discovery of FRB 20190203}

Taking into account the search criteria for dispersed pulses, one pulse has been found during the metadata analysis. Figure~\ref{fig:fig1} shows the dynamic spectrum and time profile of the detected pulse. The tail
typical for a scattered pulse is observed in the profile. For this pulse, the maximum in frequency channels appeared first at the highest receiving frequency and consistently appeared later and later at low frequencies. The total shift of the signal peak across all frequency channels was approximately 2~s, which corresponds to $DM=134.4\ \mbox{pc cm}^{-3}$. The $DM$ value collected in the Milky Way in the direction of the pulse source is equal to $20.3\ \mbox{pc cm}^{-3}$ according to the YMW16 model\footnote{www.atnf.csiro.au/research/pulsar/ymw16/} by \citeauthor{Yao2017} (\citeyear{Yao2017}). Thus, there is an excessive $DM =114.1\ \mbox{pc cm}^{-3}$ indicating the detection of the new fast radio burst (FRB). 
\begin{figure}
	\hspace{-3mm}\includegraphics[width=1.03\columnwidth]{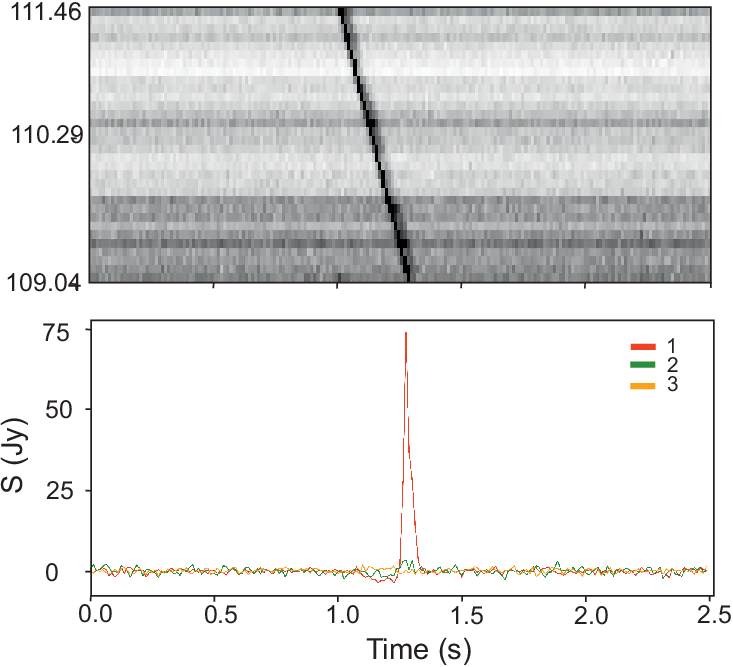}
	\caption{The same as in Figure ~\ref{fig:fig1} but the red, orange and green curves shows RRAT J1005+3015 in the main beam and the beams above and below it after compensation for $DM$. S/N for this transient reaches 234.} 
    \label{fig:fig_dop1}
\end{figure}

When summing the frequency channels without compensation for $DM$, a source with S/N=12 is visible in the raw data. In the beams above and below,  the source is not observed. After compensation for $DM$, the S/N ratio rises to 65. Such a high value of $S/N$ excludes any possibility that the registered pulse may be a statistical fluctuation, even taking into account the long-term duration of observations and the huge number of statistical trials associated with it.

Until now, the search for dispersed pulses in the PUMPS project has been carried out up to $DM \leq 100\ \mbox{pc cm}^{-3}$. The estimates by \citeauthor{Tyulbashev2022} (\citeyear{Tyulbashev2022}) show a strong signal smearing in frequency channels due to dispersion and a catastrophic deterioration in the LPA sensitivity due to scattering when searching for periodicity in pulsars on a scale of seconds at $DM \geq 200\ \mbox{pc cm}^{-3}$. In \ref{Appendix A}, we demonstrate the pulses detected from three pulsars having $DM$ significantly exceeding that of the discovered FRB. This indicates that searching for pulse signals at $DM \geq 100\ \mbox{pc cm}^{-3}$ is quite possible on the LPA3 radio telescope.

The discovered fast radio burst (FRB\,20190203) has the following parameters:
\begin{quote}
detection time (UT) 2019-02-03 02:18:10.284,\\
coordinates $\alpha_{2000}=13\uh40\um46\us\pm 2\um$,\\
\makebox[14mm]{}$\delta_{2000}=35\deg30^{\prime}\pm 10^{\prime}$,\\
dispersion measure $DM = 134.4\pm 2\ \mbox{pc cm}^{-3}$,\\ 
scattering delay  $\tau_{sc}=211\pm 6$ ms\footnote{measured from FWHM of the pulse (the galactic scattering is less than 1~ms, \citeauthor{Kuzmin2007}, \citeyear{Kuzmin2007} and will be neglected further)},\\
peak S/N $=65.$ 
\end{quote}
With the S/N ratio equal to $65$, it could be expected that the FRB will be visible in one of the neighboring beams. However there is no such signal visible at the S/N level exceeding 3. It is possible to estimate the maximum shift of the FRB coordinate in declination from the coordinate of the stationary beam, based on the beam size ($0\fdg5$) and shape ($\sin^2(x)/x^2$). If this FRB is located inside $\pm 5^\prime$ of the beam coordinate, the contribution to the neighboring beams is S/N$<3$. Thus, with a probability of about $30-35$\%, it is impossible to see the signal from the FRB in the neighboring beams, despite its high S/N.

The situation when a strong pulse is observed in a single beam of LPA3 is rare, but not exceptional. For example, in the paper by \citeauthor{Tyulbashev2018a} (\citeyear{Tyulbashev2018a}) of the 25 discovered RRATs, four had a peak flux density exceeding 20 Jy (S/N $>65$) and one of them, RRAT\,J1005+3015 (with $S_p=28.3$ Jy), has been also observed in a single beam. We processed  data for this transient obtained between August 2014 and December 2017. Its strongest pulse (see Figure~\ref{fig:fig_dop1}) had a flux density of 74 Jy and is visible in a single beam.

A two-stage calibration was carried out to determine the flux density from the found FRB. At the first step, the gain in the frequency channels was equalized using a calibration signal with a known temperature recorded in all LPA beams. In the second step, the amplitude of the found source was compared with the amplitude of the calibration source. Since the calibration source was not observed at the zenith and its coordinate did not coincide with the center of the radiation pattern, some corrections have been applied for its flux density (see some additional details in \citeauthor{Tyulbashev2019}, \citeyear{Tyulbashev2019}). To estimate $S_{p}$ we used a calibration source 4C+34.38 (J1406+34), which has a flux density of $25$ Jy. After compensation for the measured $DM$, we estimated the flux density of FRB\,20190203 to be $S_{p} = 20\pm 4$ Jy. However, the exact coordinates of the source responsible for the burst are not known. Therefore, it is impossible to correct the flux density accurately taking into account the mismatch between the centroid of the radiation pattern and coordinates of the burst. The real flux density $S_p$ can be 1.5-2 times higher.

Also, we carried out a search for repeated bursts in the direction J1340+3530, assuming different widths for a possible pulse (taking it equal to 1--32 points in extent, i.e. assuming that $W_{e}$ varies from $12.5$ to $400$ ms). We used $DM$ measured for the detected pulse and convolved the data with a model sample pulse having $\tau_{sc} = 211$ ms to obtain the maximum response (see details in the paper by \citeauthor{Brylyakova2023}, \citeyear{Brylyakova2023}). The search was carried out in the main (central) beam, where the first pulse was detected, as well as in the beams above and below it (at the right ascension $\pm 10\um$ from the coordinate measured for FRB\,190203). No new pulses with $S/N>7$ were detected in the lateral beams but a pulse with $\alpha_{2000}=13\uh39\um41\us$ was detected in the central beam in addition to the first pulse shown in Fig.~\ref{fig:fig1}. This pulse, detected in the data for June 16, 2018, has the maximum S/N$=7.2$ obtained by averaging 2 points. With further averaging, S/N begins to fall. Nevertheless, we believe that this new pulse is false (see details and discussion in \ref{Appendix B}).
\begin{figure}
	\includegraphics[width=\columnwidth]{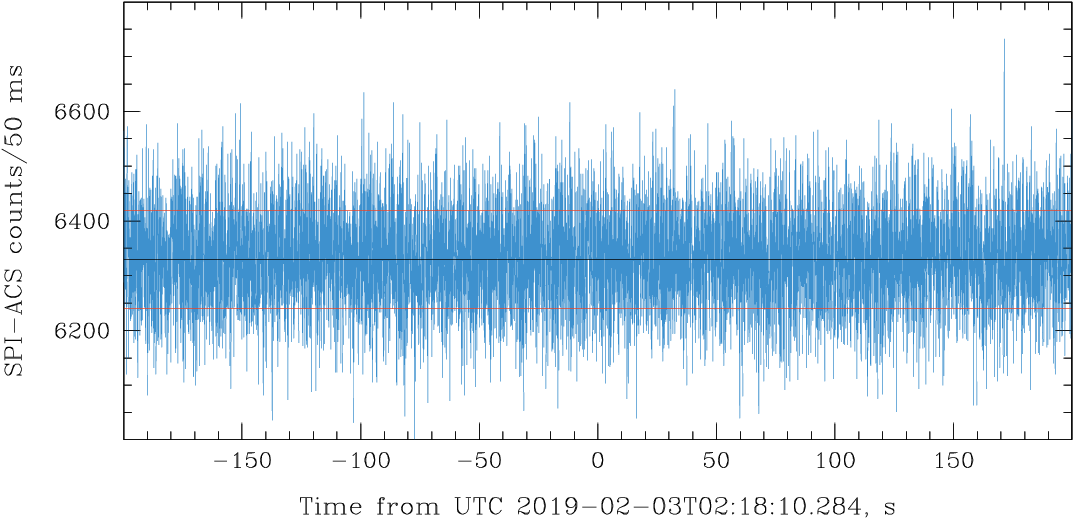}
	\caption{The SPI-ACS light curve with a resolution of 50~ms in the energy range from 80 keV to 10~MeV. The vertical scale shows the number of recorded counts. The horizontal scale shows time with $T_0 = 0$ corresponding to the LPA3 registration time of FRB\,20190203. The average background count rate is indicated by a black line, its deviations at the level of $\pm 1\ \sigma$ are shown by red lines.} 
    \label{fig:fig3}
\end{figure}

\section{Search for gamma-rays with INTEGRAL}

Detecting gamma-rays from any FRB remains a challenge. Several attempts have been made to detect gamma-ray bursts (GRBs) from FRB sources and there were even several reports on the possible detection of such GRBs (\citeauthor{DeLaunay2016}, \citeyear{DeLaunay2016}). The most reliable gamma-ray transient coinciding in both time and position with the radio burst FRB\,20200428A was detected from the soft gamma repeater SGR\,J1935+2154 (\citeauthor{Mereghetti2020}, \citeyear{Mereghetti2020}) known by its exceptional X-ray and gamma-ray activity. Recently it was assumed that FRB\,20190425A might be a counterpart of the second known gravitational wave event GW\,190425 associated with binary neutron star merger (\citeauthor{Moroianu2023}, \citeyear{Moroianu2023}). However, this hypothesis was disfavored by further considerations (\citeauthor{Bhardwaj2023}, \citeyear{Bhardwaj2023}; \citeauthor{Panther2023}, \citeyear{Panther2023}; \citeauthor{Radice2024}, \citeyear{Radice2024}; \citeauthor{Magana2024}, \citeyear{Magana2024}). The hypothesis is also challenged by the detection of GRB\,190425A with the SPI-ACS detector aboard INTEGRAL coinciding in time with GW\,190425 and believed to be its electromagnetic companion  (\citeauthor{Pozanenko2019}, \citeyear{Pozanenko2019}). There is also some contradictory evidence of transient gamma-ray radiation from the recently discovered radio burst repeater FRB\,20240114A, demonstrating extreme burst activity (\citeauthor{Xing2024}, \citeyear{Xing2024}). Note, however, that none of the five hundred FRBs detected by the Canadian Hydrogen Intensity Mapping Experiment (CHIME) were associated with any known GRBs (\citeauthor{CHIME2023}, \citeyear{CHIME2023}). Thus, any detection of the gamma-ray activity from observed FRBs would be useful for understanding the origin of FRBs and the mechanism of their emission.

The International Gamma-Ray Astrophysics Laboratory INTEGRAL (\citeauthor{Winkler2003}, \citeyear{Winkler2003}) and the Gamma-ray Burst Monitor (GBM) aboard the {\sl Fermi\/} gamma-ray observatory are the most effective experiments currently in operation for studying the cosmic GRBs and other gamma-ray transients (\citeauthor{Kuulkers2021}, \citeyear{Kuulkers2021}). On INTEGRAL, GRBs are detected within the fields of view (FOV) of its main telescopes, IBIS/ISGRI ($30\deg\times 30\deg$ FWZR) and SPI ($25\deg\times 25\deg$ FWZR) (e.g., \citeauthor{Mereghetti2003}, \citeyear{Mereghetti2003}; \citeauthor{Vianello2009}, \citeyear{Vianello2009}; \citeauthor{Minaev2014}, \citeyear{Minaev2014}; \citeauthor{Bosnjak2014}, \citeyear{Bosnjak2014}; \citeauthor{Higgins2017}, \citeyear{Higgins2017}; \citeauthor{Gotz2019}, \citeyear{Gotz2019}), but also outside their FOV (so-called off-axis GRBs) --- directly by the detectors of the gamma-ray telescopes (\citeauthor{Chelovekov2019}, \citeyear{Chelovekov2019}) or their massive anti-coincidence shield (first of all by SPI-ACS, e.g., \citeauthor{Rau2005}, \citeyear{Rau2005}; \citeauthor{Savchenko2012}, \citeyear{Savchenko2012}; \citeauthor{Minaev2014}, \citeyear{Minaev2014}; \citeauthor{Biltzinger2023}, \citeyear{Biltzinger2023}). An important advantage of the observatory is its ability for the persistent monitoring of the almost entire sky.

INTEGRAL allows one to search for transient gamma-ray radiation that could accompany the detected FRB, as well as for a possible gamma-ray source within the localization region of this FRB. The first task could be solved with the omnidirectional scintillation detector SPI-ACS. It registers gamma-ray photons in a single wide energy channel of 80~keV -- 10~MeV with a time resolution of 50~ms. The count rate is recorded continuously throughout the spacecraf's orbit (currently 68 hours), with the exception of a few hours when it passes through the radiation belts of the Earth. At the time of the FRB registration 2019-02-03 (UT) 02:18:10.284, the observatory was at a distance of 102,200 km from the center of the Earth, in the region with a rather stable radiation background. The angle between the FRB source and the X-axis of the observatory, which coincides with the main axis of the coded-aperture telescopes, was equal to 99\deg. Thus, the FRB localization region was outside FOV of the telescopes. But, it was not overshadowed by the Earth disk.

\subsection{Prompt gamma-ray burst}

Gamma-ray emission that could accompany FRB\,20190203 should be detected by SPI-ACS. However, we have not found any statistically significant excess over the background level on a time scale of 50~ms during $\pm$100~s from the time of the FRB registration. The $3\sigma$ upper limit for a short ($<50$ ms) excess in the count rate is equal to 270 counts per bin. It corresponds, after recalculation based on the SPI-ACS calibration by \citeauthor{Minaev2023} (\citeyear{Minaev2023}), to the upper limit for the 10--1000 keV fluence of a short ($<50$ ms) GRB equal to $1.2\times 10^{-7}\ \mbox{erg cm}^{-2}$ in the case when it has a soft energy spectrum (with spectral hardness $E_{p}=20$ keV) and $4.8\times 10^{-8}\ \mbox{erg cm}^{-2}$ in the case of its hard spectrum (with $E_{p}=500$ keV). The SPI-ACS light curve is shown in Fig.~\ref{fig:fig3}.
\begin{figure}
	\includegraphics[width=\columnwidth]{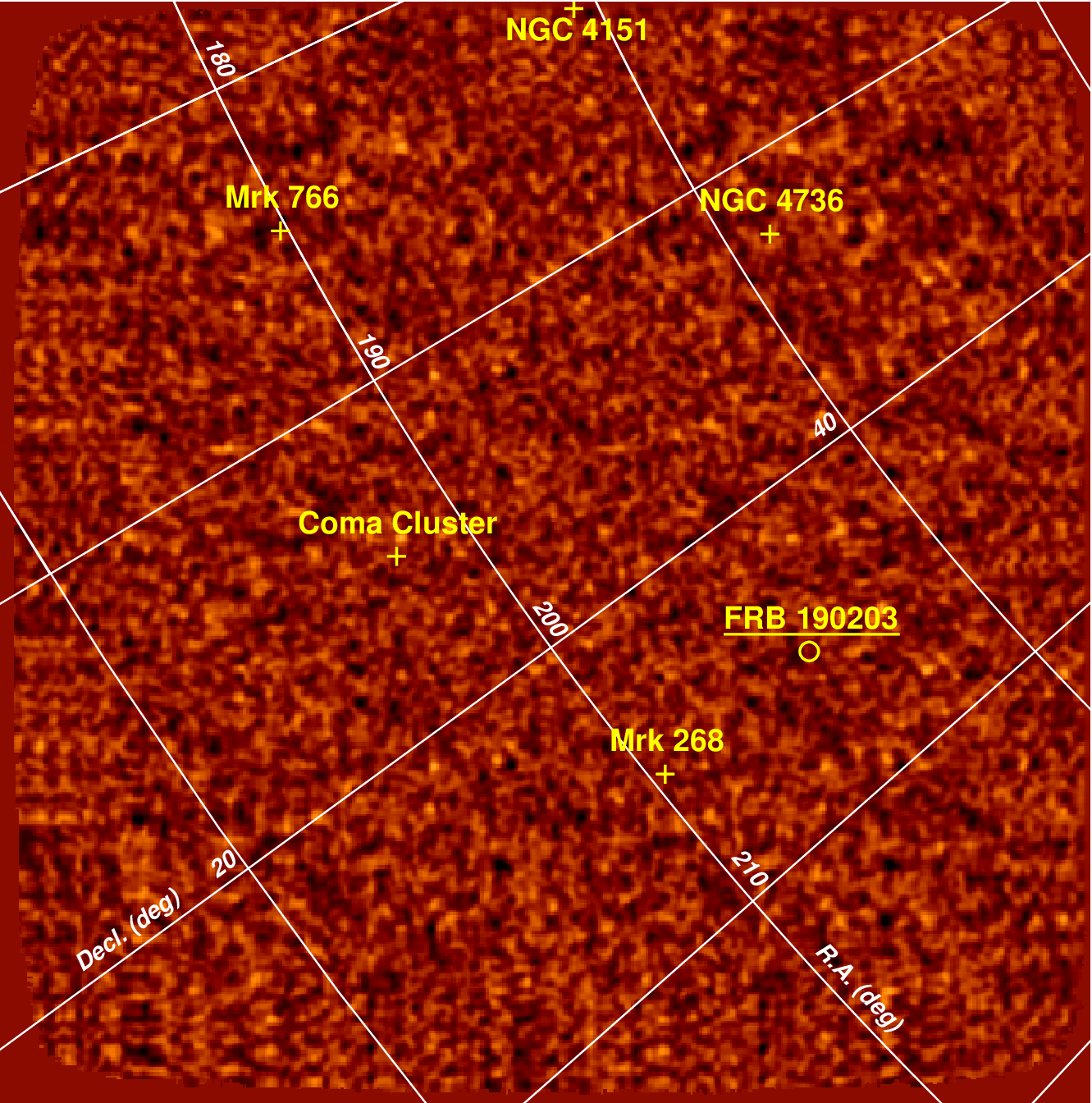}
	\caption{The S/N map of the sky area in the field of view of the INTEGRAL IBIS/ISGRI telescope obtained in the 20--150 keV range on June 10, 2005 at (UT) 22:39:35.381. It contains the localization region of FRB\,20190203. The exposure is 0.7 s. The lines on the map show right ascension and declination in degrees.}
    \label{fig:fig4}
\end{figure}

The source could have been registered by other telescopes aboard INTEGRAL. The most suitable for such out-of-aperture registration is the position sensitive detector ISGRI located in the focal plane of the IBIS gamma-ray telescope. ISGRI registers hard X- and gamma-rays in the 20--400~keV range. Note that many cosmic GRBs have been indeed recorded in this way (\citeauthor{Chelovekov2019}, \citeyear{Chelovekov2019}). However, our analysis of the IBIS/ISGRI data has not revealed any significant excess above the background level on the time scales of 5~ms, 10~ms or 50~ms during 100~s before the FRB\,20190203 registration (GRB probably should precede the radio burst because of scattering).

\subsection{Hard X-ray activity on long-term scale }

During the more than 20 years of the INTEGRAL operation on orbit, the localization region for FRB\,20190203 has repeatedly come into view of the observatory's main telescopes, including in the IBIS/ISGRI field of view. The total exposure for this localization region having been in the IBIS FOV was 73.244 days. A direct search for possible short GRB events within the FRB\,20190203 localization region with such a long exposure is very difficult, so a special technique for this analysis was developed and used (see \ref{Appendix C}).

In total, 112 candidates to bursts exceeding the selected threshold of $10 \sigma$ on a time scale of 10~ms and 15 candidates exceeding the same threshold on a scale of 50~ms have been found in the IBIS/ISGRI detector's light curves using the mentioned technique. Unfortunately, none of these candidates has been confirmed on the related sky images, none has occurred to be a real burst. For example, Fig.~\ref{fig:fig4} shows the S/N map of the sky obtained in the 20--150 keV range within the IBIS FOV for the candidate to a burst detected on the light curve on June 10, 2005 at (UT) 22:39:35.381 on a time scale of 0.01 s. The duration of the burst and the exposure used to construct the map was 0.7 s. The position of FRB\,20190203 is located close to the map center and indicated by a circle. No source is found at this position at the $S/N$ level exceeding 3, moreover there is no significant $S/N$ excesses on the entire map. Crosses indicate the positions of the Coma cluster and several other known X-ray sources (AGNs) that happen to be  present in the field. None of them appeared on the map due to the very short exposure.

The $3 \sigma$ upper limit for the flux in the 20--150 keV range from a possible source at the FRB\,20190203 position is equal to 360~mCrab ($8\times 10^{-9}$ erg cm$^{-2}$ s$^{-1}$). This limit can be considered typical for all detected ``bursts''. It is likely that all the 112+15 ``bursts'' found in the IBIS/ISGRI detector's light curve are connected with fluctuations or have geophysical origin.

\section{Discussion}
A few comments should be given regarding the discovered FRB\,20190203:

1). \citeauthor{Karastergiou2015} (\citeyear{Karastergiou2015}) carried out a search for FRBs at a frequency of 145 MHz using several LOFAR stations. No FRBs have been detected and an upper estimate given for the expected number of FRBs with a flux density of $S_p> 62$~Jy for the entire sky per day was 29 events. Our observations were carried out at a frequency of 111 MHz, which is close to 145 MHz, so we can use this upper estimate to evaluate the expected number of FRBs detectable with LPA3 per day. Up to 12 FRBs per day can flare up in the 17,000-square-degree sky area covered by the daily LPA3 survey. But, taking into account the small instantaneous field-of-view of LPA3, which is only 38 square degrees, we can expect to detect less than $12 \times (38/17,000)=0.027$ events per day or 80 FRBs for 3000 days of our observations. In fact, we detected only one event (FRB\,20190203) which suggests that the detection rate is 80 times lower than expected. 

Either the LOFAR upper estimate for the expected rate of FRBs was too optimistic (indirectly it is confirmed by the very small number of FRBs detected at low frequencies to date) or we need to attribute such an incredibly low LPA3 detection rate of FRBs at 111 MHz to the low sensitivity of the metadata search.

Another explanation may be related to the idea that FRBs emit in a narrow frequency range. Indeed, any upper limit of the total number of FRBs in the entire sky implies that FRB emission is generated over a wide frequency range, but if FRB intrinsic emission occurs in a narrow band (as the model by \citeauthor{Khangulyan2022} (\citeyear{Khangulyan2022}) suggests), then any estimates for a narrower range, such as that of LPA3, should be significantly smaller.

2). The trigger for signal detection in the metadata was the appearance of a peak with a flux density of $>7.5$ Jy in individual frequency channels, which approximately corresponds to S/N$=5$ for these channels. A trigger operating on separate channels makes it possible to isolate strong pulse interference, but is too rough to search for pulsed dispersed signals throughout the entire observation band. In a standard search for dispersed signals $DM$ is sorted out with subsequent summation of frequency channels, which in the case of 32-channel observations on LPA3 will lead to an increase in the S/N ratio by $32^{1/2} = 5.65$ times. Furthermore, scattering widens the pulse, and additional time averaging is necessary to obtain the maximum S/N value. For the discovered FRB\,20190203, averaging over 211~ms/12.5~ms$ =16$ points will increase the S/N ratio by a factor of 4. Thus, a standard search for pulsed dispersed signals in PUMPS data with sorting out of $DM$ and pulse width allows bursts that are an order of magnitude weaker than those considered in this paper to be detected. We are currently developing a new FRB search program.

3). It is known that cosmic plasma scattering can differ by an order of magnitude or greater from the average value at the same $DM$ (\citeauthor{Bhat2004}, \citeyear{Bhat2004}; \citeauthor{Kuzmin2007}, \citeyear{Kuzmin2007}). Let us estimate the expected scattering at the observed value of $DM$ for FRB\,20190203 using the formula given by \citeauthor{Cordes2003} (\citeyear{Cordes2003}):
\begin{equation} \label{eq:tau}
\begin{array}{l@{\,}l@{\,}l}
\log \tau_{sc}(\mu\mbox{\rm s})&=&-3.59+0.129\,\log DM +1.02\,(\log DM)^2\\
 & &-4.4\,\log \nu (\mbox{\rm GHz}).\\ 
\end{array}
\end{equation}
Substituting the value $DM=134.4\ \mbox{pc cm}^{-3}$ and $\nu=0.11025$ GHz, we get the expected scattering $\tau_{sc}=330$ ms. Taking into account possible deviations of the scattering by an order of magnitude from the expected value, our estimate $\tau_{sc}=211$ ms is reasonable close to the obtained value.

4). The distance to the burst can be determined by the observed excessive $DM$. We don't know the environment inside the host galaxy of the burst. If we assume that ther is no contribution from the host galaxy to the observed excessive $DM$, then according to the YMW16 model (\citeauthor{Yao2017}, \citeyear{Yao2017}) the redshift $z$ is equal to 0.17 and the luminosity distance to the burst is $d_{L}\simeq 713$ Mpc ($d_L\approx 10^{27.3}$ cm). This value is an upper limit for the distance to the discovered FRB. In further calculations, we will rely on this value, but we will leave the dependence of all observed and calculated values on it and introduce a normalization coefficient ($d_{L,27.3}\le1$). Here and further in the paper, we used the following notation in {\sc cgs} units $A_a=A/10^a$.
\begin{figure}
	\includegraphics[width=\columnwidth]{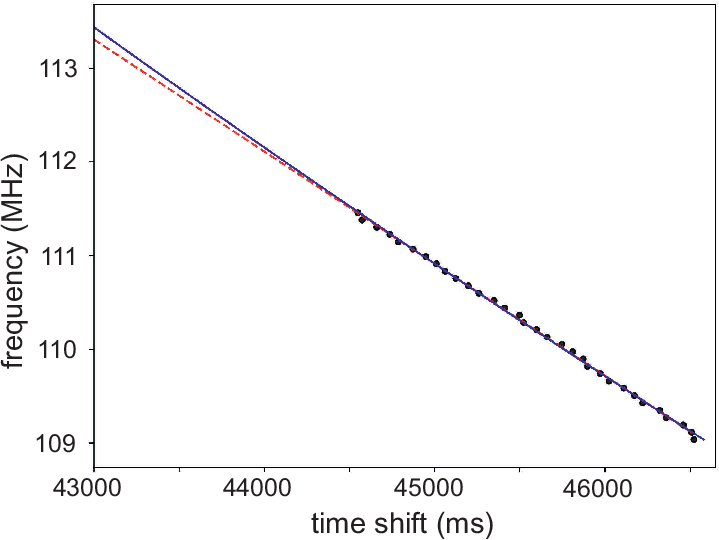}
	\caption{The figure shows the fit of the obtained data with a logarithmic straight line (red strokes) and a logarithmic parabola (blue line). There are the observation frequency on the vertical axis and the dispersion shift relative to the infinite frequency on the horizontal axis. Black points are the dispersion shift in 32 channels obtained from the dynamic spectrum according to Fig.~\ref{fig:fig1}.}
    \label{fig:fig5}
\end{figure}

5). At a known distance to the FRB, its luminosity can be estimated. We use the formula by \citeauthor{Petroff2019} (\citeyear{Petroff2019}):
\begin{equation}\label{eq:L}
    L=(4\pi d_L^{\,2}S_\nu \Delta \nu)/(1+z),
\end{equation}
where $L$  is the isotropic equivalent of the source luminosity in the frequency band $\Delta \nu$, $d_L$ is luminosity distance, determined taking into account the redshift $z$ of the source. Substituting the measured flux $S_{\nu}=20\pm4$ Jy, the channel width 2.5 MHz and $z\simeq0.17$ in Eq.~(\ref{eq:L}) we obtain
\begin{equation}\label{eq:Lum}
  L\simeq 3\times 10^{40}\ d^{\,2}_{L,27.3}\ \mbox{erg s}^{-1}.
\end{equation}
According to the analysis by \citeauthor{Cui2022} (\citeyear{Cui2022}) of the FRB luminosities detected in CHIME, the lower limit for the luminosity in the sample of 135 FRBs is at the level of $10^{40} - 10^{42}\ \mbox{erg s}^{-1}$ which is consistent with our estimate for the luminosity of FRB\,20190203 (Eq.~\ref{eq:Lum}).

At the same time, it must be remembered that the resulting estimate for the luminosity depends on the frequency band used for its evaluation. In particular, the  bandwidth of CHIME is equal to 400 MHz, which is 160 times (two orders of magnitude) wider than the LPA3 reception bandwidth. Obviously, the luminosity will be higher when observing the same burst in a wider reception band. When comparing the luminosities of bursts measured by different instruments with different receiving bands and different central frequencies, it makes sense to use the luminosity normalized per unit frequency ($\Delta\nu$). In our case, we get 
\begin{equation}  \label{eq:Lum1}
    L_\nu \simeq 10^{34} d_{L,27.3}^{\,2}\ \mbox{erg s}^{-1} \mbox{Hz}^{-1}.
\end{equation}
Figure 16 in the paper by \citeauthor{Petroff2019} (\citeyear{Petroff2019}) shows estimates of FRB normalized luminosities obtained by different telescopes except those found by CHIME. These luminosities are in the range $L_\nu \sim 10^{28} - 10^{34}$ erg s$^{-1}$ Hz$^{-1}$. The normalized luminosity of FRB\,20190203 is close to the brightest bursts in the figure.

6). A short ($\le 50$ ms) gamma-ray signal coinciding in time with FRB\,20190203 was not detected from the FRB location at energies $>80$ keV. The $3\sigma$ level for its fluence in the 10-1000 keV range was $4.8\times 10^{-8}\ \mbox{erg cm}^{-2}$. Assuming the signal duration to be $\sim 50$ ms and substituting the above distance from the FRB source, we can estimate its maximum luminosity in gamma-rays at which it can still be detected as 
\begin{equation} \label{eq:Eg}
    L_{\gamma} \leq 6\times10^{49} d_{L,27.3}^{\, 2}\ \mbox{erg s}^{-1}.
\end{equation}
Note that such a high gamma-ray luminosity corresponds, however, to a very moderate normalized luminosity in terms of Eq.~(\ref{eq:Lum1})  $L_{\gamma,\nu} \leq 2.4\times10^{31} d_{L,27.3}^{\, 2}\ \mbox{erg s}^{-1} \mbox{Hz}^{-1}$.

7). Many models have been proposed to explain the origin of FRBs (\citeauthor{Lyubarsky2014}, \citeyear{Lyubarsky2014}; \citeauthor{Metzger2019}, \citeyear{Metzger2019}; \citeauthor{Cordes2019}, \citeyear{Cordes2019}; \citeauthor{Petroff2019}, \citeyear{Petroff2019}; \citeauthor{Xiao2021}, \citeyear{Xiao2021}; \citeauthor{Zhang2022}, \citeyear{Zhang2022}). In our opinion, the observed properties of FRB\,20190203 are best explained by the model of a synchrotron maser source excited by a magnetar (\citeauthor{Khangulyan2022}, \citeyear{Khangulyan2022}).  For the normalized luminosity given by Eq.~(\ref{eq:Lum1}) and assuming that the radio pulse occupied 20 MHz in the frequency range, as measured in repeating bursts of FRB\,20180916B by LOFAR (\citeauthor{Pleunis2021}, \citeyear{Pleunis2021}), we have the total radio luminosity $L_{r} = 2\times 10^{41}\ \Delta\nu_{7.3} d^{\,2}_{L,27.3}\ \mbox{erg s}^{-1}$, here we entered a notation of $\Delta\nu_{7.3}$ for the FRB pulse width (in 20 MHz units).

We observed a scattered burst, which became significantly wider than it was at birth. The duration of almost all FRB pulses is in the range of 1--10 ms\footnote{www.astronomy.swin.edu.au/pulsar/frbcat/}$^,$\footnote{www.chime-frb.ca/catalog}$^,$\footnote{blinkverse.alkaidos.cn/} (\citeauthor{Petroff2016}, \citeyear{Petroff2016}; \citeauthor{CHIME2021}, \citeyear{CHIME2021}; \citeauthor{Xu2023}, \citeyear{Xu2023}). If the observed pulse at birth was as narrow as the pulses observed at high frequencies, the physical power of the radio emission in vicinity of its source may be $100/dt_{-2}$ times higher, where $dt_{-2}$ is a duration of FRB in centiseconds ($dt=dt_{-2}/100$). That is, the luminosity of radio emission from the FRB\,20190203 source during this pulse could reach the value
\begin{equation} \label{eq:L0}
    L_0\sim 2\times 10^{43} \, \Delta\nu_{7.3}\, d_{L,27.3}^{\, 2}\,/\,dt_{-2}\  \mbox{\rm erg s}^{-1},
\end{equation}
making the discovered event one of the most powerful FRBs known. Of course, the quadratic dependence of luminosity on distance is somewhat formal, since all estimates are given for the expected value of $z=0.17$.

Using the burst parameters, $L$ and $\nu$, it is possible to estimate the required power of a shock wave caused by a magnetar flare $\Dot{E}_{fl}$ and going to infinity (see equation 19 in the paper by \citeauthor{Khangulyan2022}, \citeyear{Khangulyan2022}):
\begin{equation}
    \Dot{E}_{fl} \approx 6\times 10^{44} \, R_{15}^{4/3}\ \Gamma_{fl,3}^{\,4/3}\ \sigma_{fl,-2}^{\, 1/3}\ 
    \mbox{\rm erg s}^{-1}
    \label{eq:Lflmin}
\end{equation}
Here $R_{15}$ is the distance from the magnetar to the front of the standing shock wave, $\sigma_{fl,-2}$ is magnetization in the flare, and $\Gamma_{fl,3}$ is the flare Lorentz factor. The radiation efficiency, that is, the degree of conversion of the flare energy into radiation, can be estimated by dividing the observed physical luminosity by the kinetic power of the flare $\epsilon_{eff}$. Combining Eqs.(\ref{eq:L0}) and (~\ref{eq:Lflmin}) we get
\begin{equation}    \label{eq:fleff}
    \epsilon_{eff}=\frac{L_0}{\Dot{E}_{fl}}\sim 0.04 \, \Delta\nu_{7.3}\, d_{L,27.3}^{\, 2}\, t_{-2}^{\,-1}\,  R_{15}^{-4/3}\, \Gamma_{fl,3}^{\,-4/3}\, \sigma_{fl,-2}^{\,-1/3}
\end{equation}
As can be seen, the degree of flare energy conversion should be high enough, at the level of a few percents, which is consistent with the parameters obtained in previous studies (\citeauthor{Khangulyan2022}, \citeyear{Khangulyan2022}; \citeauthor{Barkov2022}, \citeyear{Barkov2022}).  

8). The narrow frequency band of LPA3 does not allow us to see the parabolic behavior of the dispersion curve. Fig.~\ref{fig:fig5} shows that to distinguish the logarithmic line from a logarithmic parabola we must have a frequency band at least 4--5 MHz.\\

Note finally that all our estimates in this section were made based on the assumption that the $DM$ accumulated in the Galaxy is $20.3\ \mbox{pc cm}^{-3}$, and the $DM$ in the host galaxy is negligible. At the same time, it is known that the $DM$ in the Galaxy can be between 20 and 30 $\mbox{pc cm}^{-3}$ (see Fig.~2 in the paper by \citeauthor{Cook2023}, \citeyear{Cook2023}). If we assume $DM_{\rm host} = 100\ \mbox{pc cm}^{-3}$ (approximately 60\% of all ATNF pulsars \citeauthor{Manchester2005}, \citeyear{Manchester2005} in the Galaxy have $DM>100 \mbox{pc cm}^{-3}$ ), then $z=0.033$, $d=146$ Mpc. In this case, the burst occurred close to the Galaxy, $L$ will decrease by 25 times and the burst will become quite ordinary. We believe that only the upper estimate of the distance, which determines the maximum luminosity, is reliable. Other estimates are speculative.

\section{Conclusion}
We report the discovery of an intense ($S_{p}=20\pm4$ Jy) dispersed ($DM=134.4\pm 2\ \mbox{pc cm}^{-3}$) radio pulse at 111 MHz with the Pushchino LPA radio telescope. The pulse was detected in the direction of $\alpha = 13\uh41\um46\us$, $\delta = 35\deg30^\prime$ (epoch 2000.0, uncertainty $10^\prime$) on (UT) 2019-02-03 02:18:10.284. It lasted $211\pm6$ ms (FWHM after $DM$ correction). The large measured $DM$ indicates its extragalactic origin (it corresponds to the luminosity distance 713 Mpc after subtracting the Milky Way contribution). Further analysis has shown that the pulse is likely another fast radio burst; we called it to be FRB\,20190203. Repeated radio bursts from the source have not yet been detected. No also any activity has been detected in the gamma-ray range (according to INTEGRAL observations).

FRB\,20190203 is the first burst of extragalactic origin discovered in the PUMPS survey. It is also the second FRB detected at such a low frequency (111 MHz), and the first one among nonrepeating FRBs. This unique event is one of the most powerful FRBs detected so far.

\section*{Acknowledgement}
SAT, EAB are grateful to the Russian Science Foundation 22-12-00236\footnote{rscf.ru/project/22-12-00236/} for supporting the work in part of the FRB search with LPA. ASP, SAG, IVC, PYuM, and MVB  are grateful to the Russian Science Foundation (project no. 23-12-00220\footnote{rscf.ru/project/23-12-00220/}) for partial support of the INTEGRAL observatory data reduction and analysis. VAS and ASP were  supported in part through computational resources of HPC facilities at NRU HSE (project 663983).

\section*{Data Availability}
The PUMPS survey is not finished yet. The raw data underlying this paper will be shared on reasonable request to the corresponding author. The INTEGRAL data may be obtained through the public archive of the mission\footnote{www.isdc.unige.ch}.

\appendix

\section{Pulses detected at $DM>130$ pc/cm$^3$.}
\label{Appendix A}
\begin{figure*}
	\includegraphics[width=0.8\textwidth]{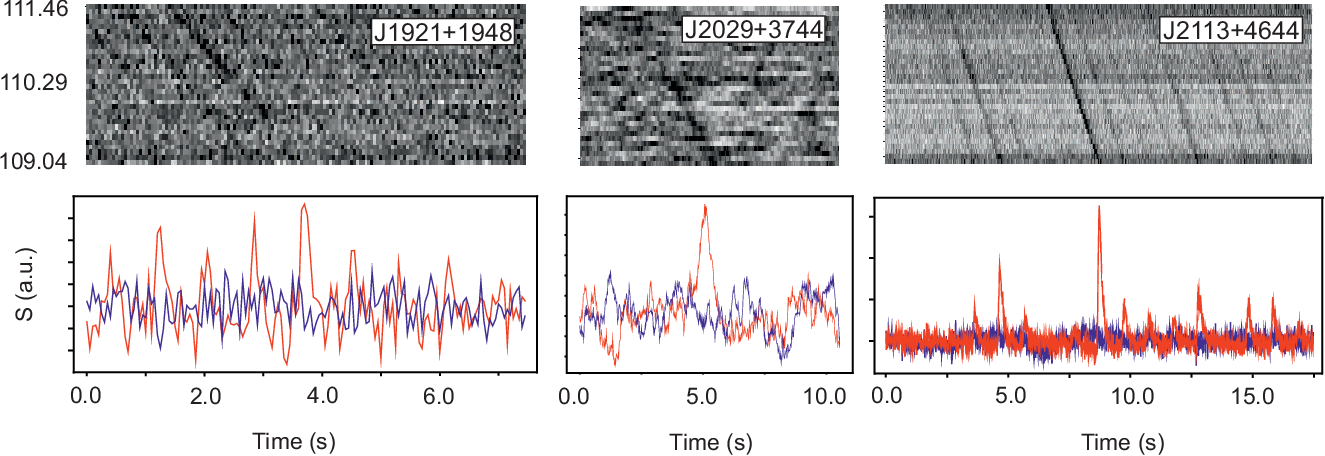}
	\caption{The upper panel shows dynamic spectra. The vertical axis is frequencies in MHz. The lower panel shows pulse profiles. The vertical axis represents the flux density in arbitrary units (a.u.). Time is shown on the horizontal axis. The time scale is the same for the upper and lower panels. The blue color is the summation of frequency channels without $DM$ compensation, and the red color is the summation of channels with $DM$ compensation.}
    \label{fig:fig6}
\end{figure*}

A blind search using summed power spectra showed that at $DM>130\ \mbox{pc cm}^{-3}$, the LPA3 radio telescope detects 10 ordinary pulsars with periods in the range of seconds  (\citeauthor{Tyulbashev2024}, \citeyear{Tyulbashev2024}). The maximum value of $DM$ was observed for the pulsar J1950+3001 ($P=2.7889$ s, $DM=249~\mbox{pc cm}^{-3}$). To search for pulsed radiation from these 10 pulsars, we selected 4 of the strongest (J1921+1948, J2029+3744, J2113+4644, and J1935+1616), the S/N ratios for harmonics in the summed power spectra of which ranged from 40 to 250\footnote{https://bsa-analytics.prao.ru/en/pulsars/}. When searching for pulsed radiation, dispersed pulses were found in 3 of the 4 pulsars (see Fig.~\ref{fig:fig6}). Pulses from J2113+4644 were found in a standard search for dispersed pulses from rotating radio transients (RRAT) having a value of $100<DM<200\ \mbox{pc cm}^{-3}$. 

The profiles on the bottom panel were obtained with different processing of data. For pulsar J2113+4644 (B2111+46: $P=1.0146$ s, $DM=141.2$ $\mbox{pc cm}^{-3}$, $W_e=149$ ms), in the lower right corner, a simple summation of channels with $DM$ compensation is carried out. On the upper (dynamic spectrum) and lower (profiles) parts of the figure, more than half of the pulses are visible, and the scattering of strong pulses is also clearly visible. For the pulsar J1921+1948 (B1918+19: $P=0.8210$ s, $DM=153.8\  \mbox{pc cm}^{-3}$, $W_e=175$ ms) averaging by 4 points was carried out, which doubled the S/N and allowed us to see the pulse profiles. Almost all of its pulses are visible for this pulsar. The pulsar J2029+3744 (B2027+37: $P=1.2168$ s, $DM=190.6 \mbox{pc cm}^{-3}$) is shown in the middle part of Fig.~\ref{fig:fig6}. With the usual averaging of points, it is poorly visible. Based on it's $DM$ value, we got a model pulse (see details in \citeauthor{Brylyakova2023}, \citeyear{Brylyakova2023}) and carried out convolution in each frequency channel independently. Thus, the convolution of the detected pulse with the model pulse is shown in the dynamic spectrum of the pulsar for each frequency channel. To obtain the profile, we performed the summation of channels taking into account the known $DM$ value of the pulsar. If for the pulsars J1921+1948 and J2113+4644 pulses were observed during hundreds of days, for the pulsar J2029+3744, only one pulse with a half-width $W_e=418$ ms was found during 3,000 observation sessions. For the pulsar J1935+1616 (B1933+16: $P=0.3587$ s, $DM=158.5\ \mbox{pc cm}^{-3}$), having S/N$ =95$ in the first harmonic in the summed up power spectrum, pulses with S/N $>7$ were not detected. It is possible that the time of scattering in the direction to the pulsar is higher than its period and the pulses merge.

Thus, we detect the pulses from at least 3 known pulsars. Their values of $DM>130\  \mbox{pc cm}^{-3}$ and exceed $DM$ of the found FRB\,20190203, which indicates a good opportunity to search for pulses from pulsars and FRBs on LPA3 with $DM$ up to $200\ \mbox{pc cm}^{-3}$.

\section{Checking the pulse found in the direction to J1340+3530}
\label{Appendix B}

\begin{figure}
	\includegraphics[width=0.8\columnwidth]{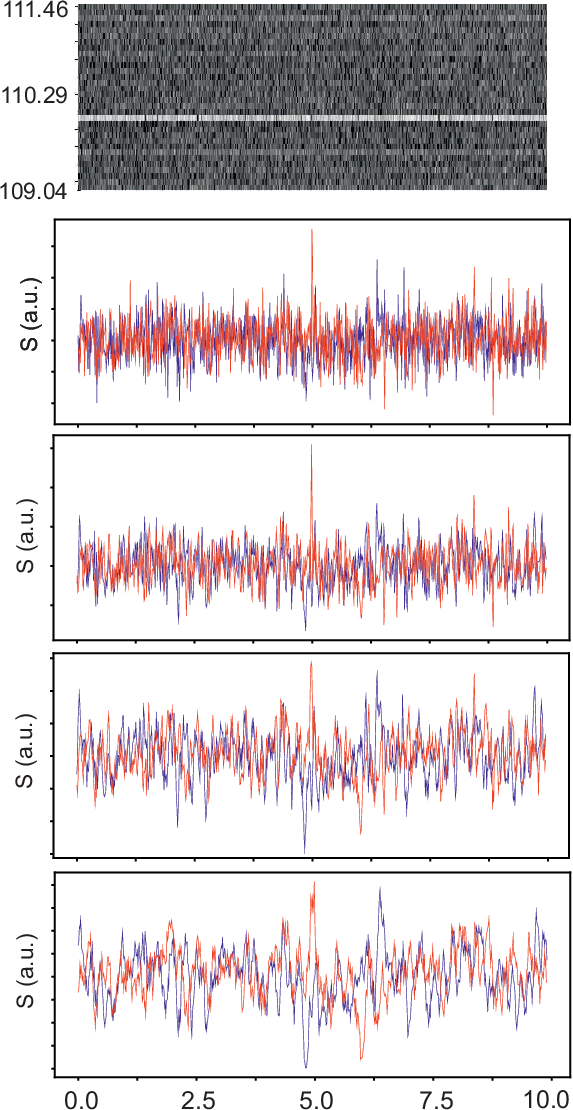}
	\caption{The upper panel shows the dynamic spectrum of the discovered pulse. Some frequency channels in MHz are marked along the vertical axis. The next 4 panels are the pulse profiles with averaging of data on 1, 2, 4, 8 points. The maximum S/N reaches when averaging by 2 points. Along the vertical axis is the flux density in arbitrary units (a.u.).  Time in seconds is shown on the horizontal axis. The time scale is given in seconds and is the same for all panels. Blue and red colors show data before and after $DM$ compensation.}
    \label{fig:fig7}
\end{figure}

Dispersion smoothing at $DM=134.4\ \mbox{pc cm}^{-3}$ in a frequency channel with a width of 78 kHz is equal to 65 ms or 5--6 points taking into account sampling equal to 12.5~ms. That is, in our observations on LPA3, the pulse having the specified $DM$ cannot be narrower than 5 points. Our estimate for the scattering of the discovered FRB\,20190203 shows that the pulse should be wider than $211/12.5=16-17$ points. This means that if we have found a real pulse, then when averaging raw data, its S/N should grow more then 2-4 times. This is not observed in Fig.~\ref{fig:fig7}. The dispersion delay line is also not visible on the dynamic spectrum. At S/N $>7$, such lines are usually visible. Therefore, we believe that the found ``repeated'' pulse is false. We were not able to find errors in processing or a source of interference that gives large dispersion delays. At the same time, the found pulse with such a value of $DM$ cannot have an extraterrestrial origin.

\section{Searching for short GRBs in the IBIS/ISGRI long-term observations}
\label{Appendix C}

To search for possible short GRBs from the FRB\,20190203 localization region, during a number of episodes when it was observed by the IBIS gamma-ray telescope, we developed a special technique based on the analysis of the IBIS/ISGRI detector's count rate history. The technique reduces the computation time required for analysis and minimizes the likelihood of a false GRB detection. The following analysis of archived data of IBIS/ISGRI has been carried out:\\ 

\begin{itemize}
\item[-]  At first, the standard programs from the basic package for INTEGRAL data processing (OSA version 10.2) have been applied to analyze the related IBIS/ISGRI data up to the DEAD level;

\item[-] Then we have applied the ``evts\_extract'' program of the OSA package to receive a list of events (registered photons) adjusted for the energy and ``dead'' time of the detector. The list does not include events from the ISGRI ``hot'' pixels. The Good Time Intervals (GTI) optimal for the analysis have been selected. For each photon from this list, there was information about the light illumination of the corresponding detector pixel (in which the photon was registered) by this source of FRB (Pixel Illumination Factor or PIF);

\item[-] Then we have constructed a light curve (count rate history) of the source in the range of 20-100~keV using events from the list with PIF<0.1 to renormalize it and use as background for the light curves with PIF$>0.2$ and PIF$>0.4$. The light curves with time bin size of 0.01 and 0.05 s have been obtained. This procedure allowed us to clear these light curves from any flashes of geophysical origin illuminating the entire detector;

\item[-] Each of these light curves has been divided into segments containing 5,000 bins, the average value and sample variance have been calculated for each segment, and all significant (over 10 standard deviations) enhancements in the count rate were revealed and collected; 

\item[-] Then the whole procedure has been repeated for the light curves divided into the new segments with an offset of 1,667 bins (to ensure that all the enhancements found near the segment edges are valid). For each of the detected enhancements (candidates to bursts), a map of the detector has been obtained, using only the photons registered during the enhancements, and compared with the map obtained during the entire observation session. If the proportion of photons from pixels containing more than 2 photons exceeded 20\%, such a burst was assumed ``false'', since it was more likely associated with charged particles rather than X-ray photons;

\item[-] Images of the gamma-ray sky within the IBIS field of view have been constructed for all the remaining candidates to bursts, using a coding aperture of the telescope, and then it was compared with the images obtained during the previous time intervals of the same duration in order to confirm the burst confidence and identify the burst source. An original program developed at the Space Research Institute was used to construct images at such a short time scale, its general principles were described by \citeauthor{Revnivtsev2004} (\citeyear{Revnivtsev2004}).
\end{itemize}

\end{document}